# The MRO-accompanied modes of Re-implantation into $SiO_2$-host matrix: XPS and DFT based scenarios

A.F. Zatsepin[1], D.A. Zatsepin[1,2], D.W. Boukhvalov[3,4], N.V. Gavrilov[5], V.Ya. Shur[6], A.A. Esin[6]

[1]*Institute of Physics and Technology, Ural Federal University, Mira Str. 19, 620002 Yekaterinburg, Russia*
[2]*M.N. Miheev Institute of Metal Physics of Ural Branch of Russian Academy of Sciences, 18 Kovalevskoj Str., 620990 Yekaterinburg, Russia*
[3]*Department of Chemistry, Hanyang University, 17 Haengdang-dong, Seongdong-gu, Seoul 133-791, Korea*
[4]*Theoretical Physics and Applied Mathematics Department, Ural Federal University, Mira Street 19, 620002 Yekaterinburg, Russia*
[5]*Institute of Electrophysics, Russian Academy of Sciences-Ural Division, 620016 Yekaterinburg, Russia*
[6]*Institute of Natural Sciences, Ural Federal University, 51 Lenin Ave, 620000 Yekaterinburg, Russia*

**Abstract**

The following scenarios of Re-embedding into $SiO_2$-host by pulsed Re-implantation were derived and discussed after XPS-and-DFT electronic structure qualification: (i) low Re-impurity concentration mode → the formation of combined substitutional and interstitial impurities with $Re_2O_7$-like atomic and electronic structures in the vicinity of oxygen vacancies; (ii) high Re-impurity concentration mode → the fabrication of interstitial Re-metal clusters with the accompanied formation of $ReO_2$-like atomic structures and (iii) an intermediate transient mode with Re-impurity concentration increase, when the precursors of interstitial defect clusters are appeared and growing in the host-matrix structure occur. An amplification regime of Re-metal contribution majority to the final Valence Band structure was found as one of the sequences of intermediate transient mode. It was shown that most of the qualified and discussed modes were accompanied by the MRO (middle range ordering) distortions in the initial oxygen subnetwork of the *a*-$SiO_2$ host-matrix because of the appeared mixed defect configurations.

# 1. Introduction

Known since ancient times, amorphous silicon dioxide ($a$-SiO$_2$) still remains of great importance for various high-tech applications because of the powerful technological potential. Having good chemical stability in the wide range of temperatures and relatively aggressive ambients, it was deeply studied since the beginning of the 20$^{th}$ century, causing the successful yield of cheap and, simultaneously, effective commercial and semi-commercial synthesis approaches ("wet", "dry" or combined "wet-and-dry" technology [1], several variations of CVD [2], aerogel [3], nanosized synthesis [4], etc.). This technologically important material has found also an extended functionality due to applied electronic structure engineering trends and controlled management of the needed functional properties via atomic structure conversion mechanisms [5-7]. Advisability of $a$-SiO$_2$ as a material for managed atomic structure conversion is determined by the peculiarities of $\equiv$ Si – O$^*$ chemical bonding unit. This major silicon-oxygen bonding unit arises due to $\sigma$-coupling of Si $s$- and Si $p$-electrons with oxygen $p$-electrons, but, as a unique specificity, the short and elonged (O 2$p$ – Si 3$d$)$_\pi$ conjugations are possible. Consequently, it is potentially giving the most wide set of dihedral angles as well as the variations in silicon-oxygen bonding distance with a critical border-values of 1.54 Å and 1.81 Å [8]. As a result, a lot of silica-oxygen polymorphs with dissimilar physical-chemical properties exist (see e.g. Ref. [6]), providing favorable conditions for the further elaboration within Research-and-Development area of SiO$_2$-based functional materials.

Despite of the powerful technological suit for electronic structure functionalization of $a$-SiO$_2$ via conventional embedding of metal or metal-like particles into Zachariasen-Warren three-dimensional glassy network, the direct ion-doping by ion-implantation techniques was proved to be promising in terms of enhanced functional performance of ion-beam treated material [9-10]. Herewith, probably, almost all the technologically widespread metals of the Periodic Table of Elements have been employed for that except rhenium – the VII group third-row transition metal. In a large variety of the most common oxidation states (+2; +4; +6; +7), usually it is exhibiting the perovskite-like atomic structure but, sometimes, an unconventional for this type of material the molecular-like unit-cells with alongside radicals. The latter depends on the extensivity of Re–Re bonding, which appears as a specific feature of the oxidation states that are lower than Re$^{+7}$ [11]. Is this a signature of concrete synthesis for an oxidation of Re-metal (some rhenium oxide specific phases are easily reduced back to metal already at ~ 400 – 500 ºC) or the sequence of vastly complex Re multi-shell many-electron configuration or both, still remains unclear. With that, the technological importance of rhenium and its compounds, that are already employed in hi-temperature combustion chambers of jet-engines, exhaust nozzles and jet turbine blades, etc., cannot be overestimated. So the mechanisms and origin of peculiarities in rhenium-oxygen electronic states interactions seems a challenging research task.

In the current paper we are presenting the results of our X-ray Photoelectron study of Re-implanted $a$-SiO$_2$ host-matrix. As it was shown earlier (see e.g. Ref. [12]), it is possible to oxidize embedded into oxide-host metal-particles by means of ion-implantation, depending on the employed synthesis mode and posterior tempering. An application of hi-purity $a$-SiO$_2$ as a "soft" (in other words self-organized) host-matrix will allow studying the processes of rhenium-oxygen interactions due to easy atomic structure re-arrangements in the used amorphous host-oxide. This will be made on the basis of detailed XPS electronic structure mapping as well as on the theoretically based discussions of experimental data obtained and the onward applied DFT-modeling. The comparison of calculated formation energies, electronic structure of valence bands (VB) and calculated densities of states (DOSes) will provide the knowledge about local atomic microstructure of the host-matrix; the distribution of impurity patterns over the volume of implanted host-matrix, stability of different atomic configurations of impurities and role of self-defects such as oxygen vacancies.

## 2. Samples, Experimental and Calculation Details

KUVI-SiO$_2$ glass (type IV) was used as a host-matrix for pulsed Re-implantation. This type of glass means conventional high-frequency plasma oxidation of appropriate initial components using optical grade standards and, as a result, this yields the high-transmission abilities in the visible, UV- and IR-regions, an enhanced radiation resistance and overall concentration of alien impurities less than $10^{-3}$ mass. %. Spectrosil™ WF (Great Britain), Suprasil™ W (Germany) and Corning 7943™ (USA) are the closest technological analogues. KUVI-SiO$_2$ matrices were prepared for implantation by standard and technologically certified clean scribing cut-off into plane-parallel plates with a surface of an optical quality, measuring $1 \times 1$ cm$^2$ and thickness of 2 mm. The geometrical parameters of the samples were chosen in order to provide their most reliable fixing in the sample-holder during Re-implantation and our onward XPS experiments.

The metal-vapor vacuum arc ion source (MEVVA-type) was employed for rhenium ion-beam obtaining. Re-cathode (as a source of Re ion-beam) was manufactured from rhenium powder of 99.9 weight % purity by pulsed magnetic pressing at 425°C. Re ion-beam was generated in a pulse-repetitive mode with 0.4 ms pulse duration and pulse repetition frequency of 25 Hz. The 30 keV Re-ions were separated for implanting from the overall beam and pulsed-repetitive ion-beam current density has been strictly limited at 0.75 mA/cm$^2$. The maximum Re-treatment time was not more than 3 h. We applied three types of Re-ion-beam treatment modes for our KUVI-SiO$_2$ host-samples: $5\times10^{17}$ cm$^{-2}$, $1\times10^{17}$ cm$^{-2}$ and $5\times10^{16}$ cm$^{-2}$. The temperature of these samples during overall Re-treatment procedures had been strictly fixed at 100 $^o$C. The oil-free vacuum pressure in the implantation chamber was held up at 0.01 Pa.

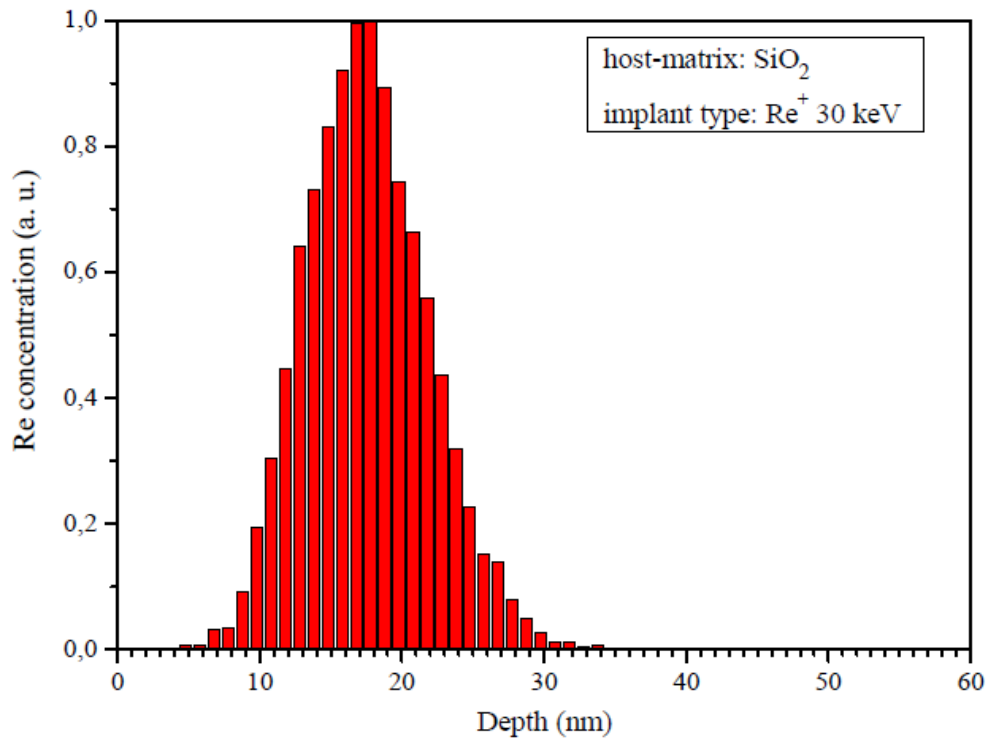

**Figure 1.** SRIM simulation of the penetration depths for 30 keV Re-ions in SiO$_2$ host-matrix.

Figure 1 displays the results of SRIM Software [13] estimation of the penetration depth profile for the 30 keV Re-ions in SiO$_2$ host-matrix. According to the known ~ 8 nm depth resolution of the XPS technique, this method will cover the beginning of the depth distribution profile estimated by SRIM calculations. Nevertheless, the XPS Thermo Scientific™ *K*-Alpha+™, having the element concentration detection limit ~ 0.05 at.% (as declared by the Manufacturer) will successfully allow to obtain the core-level and valence-band XPS Data for electronic structure mapping. Also we have to note that the accuracy of SRIM estimations for heavy ions (Z > 3) is within the range of 10% thus giving larger errors for heavy ions than that for the light ones (~ 5%) [13] with all the ensuing consequences.

Thermo Scientific™ *K*-Alpha+™ X-ray photoelectron spectrometer was used for fast wide-scan chemical contamination analysis (survey XPS spectroscopy) and onward core-level and valence band (VB) structure XPS mapping. Feature of wide-gap insulators XPS, which includes KUVI-SiO$_2$, is extremely strong differential charging of the samples due to the loss of photoelectrons and, as a result, an incorrectly distinguished binding energy chemical shift occurs in the recorded XPS spectra. As a limiting case, sometimes it is even impossible to distinguish it at all. Thus in order to overcome these stumbling blocks of the built-in XPS charge-neutralizer, well-conducting noble metal overlayer (viz. Au, Pt, Pd, ....) is implemented into the wide-gap insulating sample or 2-3 nm graphite film is applied [14]. Comparing with other conventional XPS systems, the most unique advantage of XPS Thermo Scientific™ *K*-Alpha+™ for our case is dual-beam flood gun, coupling low-energy ion-beam with very-low energy co-axial electrons (less than 10 eV), which prevents sample charging during analysis (GB Patent 2411763). This eliminates, in most cases, the need of charge referencing procedure and makes the analysis of the XPS data from insulating samples as easy as possible. Nevertheless, an extended accuracy is of high importance for the mentioned case of XPS analysis. Thermo Scientific™ *K*-Alpha+™ system is equipped with 180º double-focusing hemispherical energy analyzer and provides an energy resolution not worth than 0.28 eV. All the measurements were carried out under X-ray Al *K*α monochromatic X-ray source in an oil-free vacuum at $5 \times 10^{-6}$ Pa pressure (provided by two 250 l/s turbo-molecular pumps), 300 μm diameter of X-ray Al *K*α spot and 72 W X-ray power load of the sample, analyzer 200 eV pass-energy window in a fast-scan mode (XPS survey) and 50 eV for core-level and VB mapping mode, multi-point spectra acquisition (128-channel detector) with their posterior summing and averaging (as it was firstly implemented as advantage and justified in our previous experiment [7]). We have to note that the combined rhenium-metal and rhenium-oxygen XPS external standard was employed because of the relatively small binding energy difference (about 0.8 eV) between Re-metal and the lowest rhenium oxide phase – $ReO_2$. In some highly specific cases (viz. Zn 2p core-level spectroscopy of Zn-metal and ZnO, BE-difference ~ 0.2 eV) it is not possible at all to separate metal phase and oxide phase for this reason. In our case we have an amorphous host-matrix so the differential charging of the sample under XPS study might bring additional problems for separating the contribution from embedded Re-metal and it oxide phases. That's why the employment of a combined rhenium-metal and rhenium-oxygen XPS external standard seems to us the reasoned decision.

The results of XPS survey chemical contamination analysis are shown at Fig. 2. It is evidently seen from Figure 2 that there are no alien impurities except XPS core-level signals from appropriate electronic states of Re that had been embedded by ion-implantation into class IV KUVI-$SiO_2$ hi-pure host-matrix. C 1s core-level signal is lower than other XPS signals from electronic structure of Re-implanted samples and is totally absent in the spectrum of oleophobic and untreated KUVI-$SiO_2$ host-matrix (XPS reference sample). The reasons for that as well as the calibration and identification methodology were reported in our previous study of amorphous materials using ULVAC-PHI Versaprobe 500 XPS spectrometer [7]. Finally, the performed XPS chemical qualification analysis allows to conclude that the declared empirical formulas of the implanted samples are valid and no extra contaminations appeared after Re-ion-implantation.

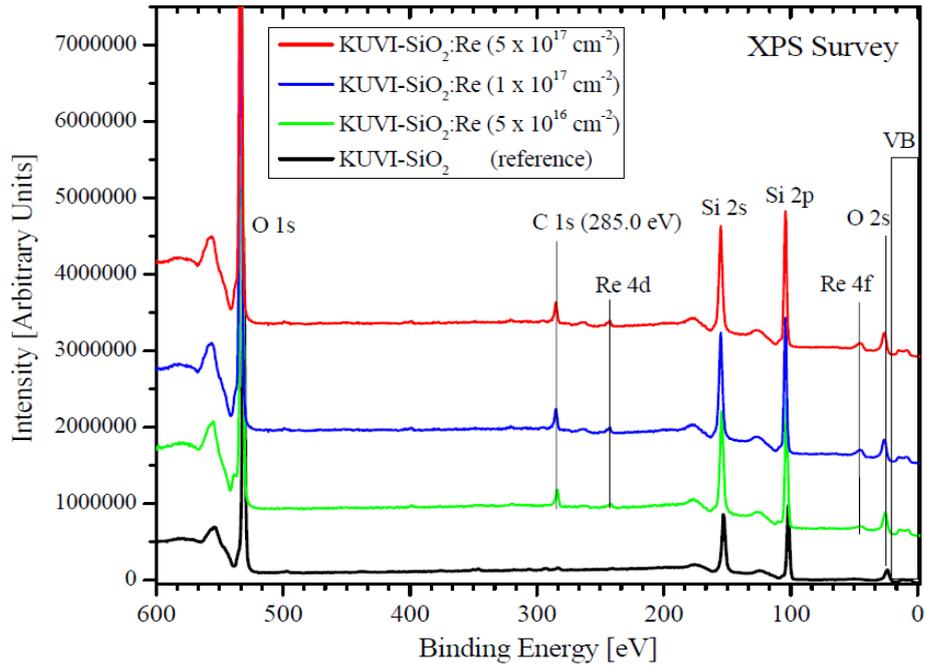

**Figure 2.** XPS Survey spectra of *as-implanted* KUVI-SiO$_2$:Re samples at $5\times10^{17}$ cm$^{-2}$, $1\times10^{17}$ cm$^{-2}$ and $5\times10^{16}$ cm$^{-2}$ fluences. The spectrum of *as-sintered* hi-pure KUVI-SiO$_2$ (reference host-matrix) is shown for comparison.

By way of theoretical base for Re-embedding scenarios the Density Functional Theory (DFT) was selected and employed. Under this approach the calculations of formation energies for different atomic configurations of structural defects induced by Re-ion implantation in *a*-SiO$_2$ were made. These calculations were performed using of the SIESTA pseudopotential code [15], a technique that has been recently successful in related studies of impurities in SiO$_2$ [12]. All calculations were made employing the generalized gradient approximation (GGA-PBE) [16] for the exchange-correlation potential in a spin-polarized mode. The fundamentals are based on the finite range pseudoatomic orbitals (PAOs) of the Sankey–Niklewsky type [17, generalized to include multiple-zeta decays. The interactions between electrons and core ions are simulated with the separable Troullier–Martins [18] norm-conserving pseudopotentials. Atomic pseudopotentials were generated separately for atoms Re, Si and O by using the $6s^2 6p^0 5d^5$, $3s^2 3p^4$ and $2s^2 2p^4$ atomic configurations, respectively. The cut-off radii for the present atomic pseudopotentials are taken as s: 2.66, p: 2.72, d: 2.47, f: 2.47 a.u. of Re, s: 1.77, p: 1.96, d and f: 2.11 and 1.45 a.u. for the s, p, d and f channels of O. The same pseudopotential was previously employed for the calculations of structural properties of rhenium nitrides. [19]

A full optimization of the atomic positions had been carried out during which the electronic ground state was consistently found using norm-conserving pseudopotentials for the cores and a double-ξ plus polarization basis of localized orbitals for Si, Re and O spices. The forces and total energies were optimized with the accuracy of 0.04 eV Å$^{-1}$ and 1.0 meV, respectively. The calculations of formation energies ($E_{form}$) were performed by considering the supercells both with and without a given defect [12]. For the current modeling of Re impurity interactions with quartz-like matrix we employed the specially build $Si_{24}O_{48}$ supercell of α-quartz. The presence of oxygen vacancies in the vicinity of all studied defects was also taken into account for considerations. While the experimental data (see Fig. 1, low signal area) demonstrates negligible amount of surface and subsurface layer impurities (i.e. below 1nm from the surface), so we will consider only the case of impurities in the bulk $SiO_2$.

In order to check of the role of spin-orbit coupling we also have performed the calculations of the electronic structure and formation energy in the collinear and non-collinear modes for the case of single substitutional impurity. It was established that the transformations of the electronic structure and formation energies are nearly negligible. This result is in a good agreement with calculations of structural parameters of rhenium nitrides where the spin-orbit coupling, being taken into account, provides the changes of lattice parameters below 0.01 Å. As for the electronic structure, here the only tiny splitting of few energy bands without any changes in bandgap were found [20]. Therefore in according with recent calculations of Re-impurities in the matrix of sp-elements [21] for decreasing of computational costs we should not take into account the spin-orbit coupling for our further modeling.

## 3. Results and Discussion

The measurements of X-ray Photoelectron core-level Re $4f_{7/2-5/2}$ spectra (Fig. 2) for *as-implanted* the KUVI-$SiO_2$:Re samples demonstrate that these spectra has a complex structure opposite to well-known double-peak core-level XPS spectra for pure metals and simple oxides, evidencing either about multi-phase composition of the Re-ion beam treated KUVI-$SiO_2$ sample, or about the XPS shake-satellites manifestation. The latter had been well XPS-recognized previously as a spectral signature for appropriate metal core-level spectra of divalent copper-oxygen compounds, allowing to employ them as an identification of concrete electronic states of an oxidized metal due to their specific coupling with partial electronic states of oxygen [22, 23]. Taking into account the multi-shell many-electron configuration of Re-atom, it will be not surprising to uncover these XPS shake-satellites for rhenium-oxygen compounds as in the case of CuO-based materials, but comparison with a combined Re metal-and-oxide XPS standard (Fig.3, red spectrum) forces us to think that otherwise is essential. This comparison allows us to reveal, that the XPS features in KUVI-SiO$_2$:Re XPS spectrum, that are located at ~ 40.1 eV and ~ 43.0 eV, belong to Re-metal 4f$_{7/2-5/2}$ core-level components; the ~ 41.8 eV and ~ 44.0 eV are well coinciding with Re 4f$_{7/2-5/2}$ of quadrivalent rhenium oxide ReO$_2$ and the low-intensity ~ 46.0 eV – 49.0 eV bands belong to that from the highest rhenium valency oxide – Re$_2$O$_7$. Leastwise all the XPS features that are forming the shape-structure of XPS Re 4f$_{7/2-5/2}$ of the sample under study are in a good agreement with reference to binding energy values of rhenium metal-and-oxide combined XPS external standard (reference). Finally, the visual spectral gravity center is nearly in the same binding energy region as for applied XPS standard thus evidencing the majority of Re-metal contribution to Re 4f$_{7/2-5/2}$ signal. This Re-metal majority might be regarded as an oversaturation of KUVI-SiO$_2$ within concrete 5×10$^{17}$ cm$^{-2}$ implantation mode (Re-loss effect) and, with no doubt, will cause the strongest transformation in valence band (VB) structure at the $E_F$ vicinity. From atomic structure point of view, this Re-oversaturation results in the interstitial accumulating of Re-metal particles or/and partial charge compensation of intrinsic non-bridging oxygen atoms in KUVI-SiO$_2$ host.

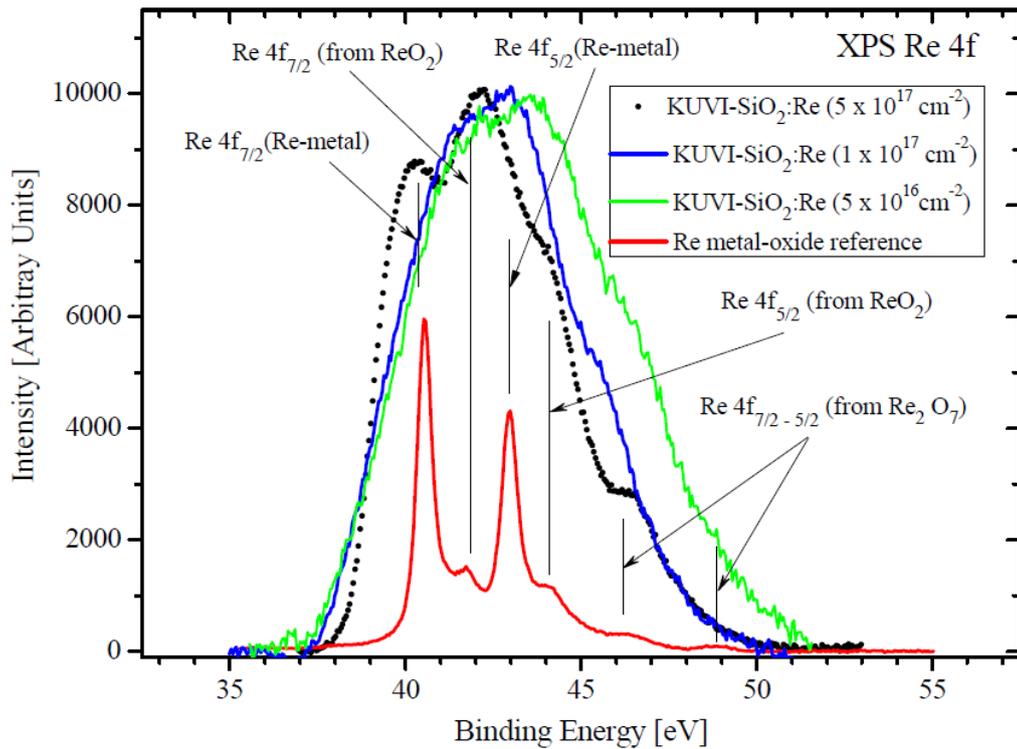

**Figure 3**. Normalized XPS Re 4f core-level spectra for Re-implanted KUVI-SiO$_2$ host-matrix after different implantation fluences without tempering.

Despite of the analyzer 50 eV pass-energy window applied in our XPS core-level analysis (this is the smallest value which is possible to apply at Thermo Scientific™ *K*-Alpha+™ without critical fall-down in signal-to-noise ratio), the Re $4f_{7/2-5/2}$ spectrum of KUVI-SiO$_2$:Re exhibits much wider base-bands than that for rhenium metal-and-oxide XPS external standard. This doesn't seem overweighted for above discussed, because, on the one hand, initially amorphous KUVI-SiO$_2$ host-matrix was applied for Re-implantation, and on the other, the spectrum of the sample with the highest fluence from the used is under discussion. Ion-implantation treatment might distort the long-range ordering in crystalline materials and even transform it from highly-ordered crystal phase to amorphous one [9-10]. For our particular case, the broadened base-band of the core-level Re $4f_{7/2-5/2}$ spectrum means an increased continual disordering in KUVI-SiO$_2$ silica-oxygen network after Re-implantation, which, from the very beginning, has more unified and both short- and middle-range ordering structure as compare with convenient *a*-SiO$_2$ (MRO-mode glass, as it was firstly named in Ref. [8]). If our supposition on the MRO-distortions in KUVI-SiO$_2$ host-matrix is valid, one will see the base-band broadening and line-shape transformations in the O 1s and Si 2p core-level spectra of the samples under study and this broadening have to be dependent on the value of implantation fluence, as it was clearly shown by Hashimoto et.al. (see Fig. 1 in Ref. [24]) and posterior tempering, if any. Besides the MRO-broadening, also the concentration of Re-metal and its oxide phases will be organically dissimilar in the used host-matrix and might be uniquely detected by XPS. But firstly we will discuss the core-level fluence dependence during Re-implantation.

The other XPS Re $4f_{7/2-5/2}$ core-level spectra for Re-implanted KUVI-SiO$_2$ host-matrix after different implantation fluences without tempering are presented at Fig. 2 as well. An intensity of rhenium-oxygen XPS external standard spectrum is suppressed by factor of 1.5 for clarity of other spectra. It is notable that the decreasing of implantation fluence from $5\times10^{17}$ cm$^{-2}$ to $5\times10^{16}$ cm$^{-2}$ results in essential band-shape transformation with the shift of spectral gravity center to the higher binding energy region (compare the black and green spectra shown on Fig. 3), viz. to the region, where rhenium oxide phases are dominating. This means that the mentioned decreasing suppresses Re-metal accumulation in interstitials and activates the solid-state interaction processes with more effective coupling of embedded rhenium atoms with oxygen atoms in KUVI-SiO$_2$ host, increasing the contribution majority of rhenium oxide phases. Let us now analyze the XPS O 1s core-level spectra of the samples under study. These spectra are shown at Fig. 4.

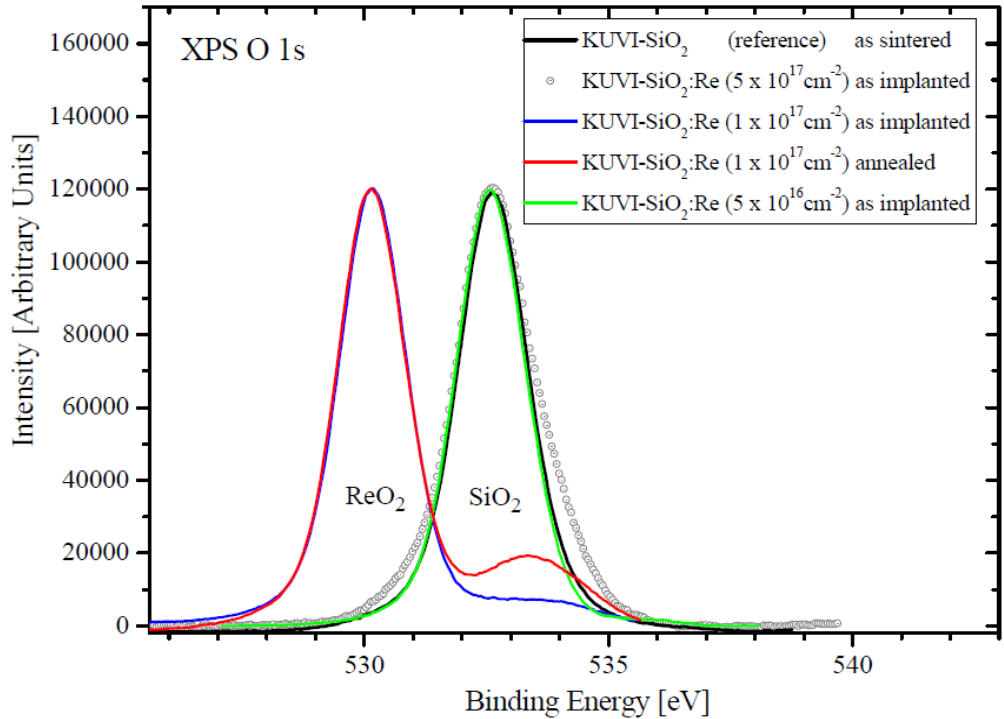

**Figure 4**. Normalized XPS O 1s core-level spectra for Re-as-implanted KUVI-SiO$_2$ host-matrix after different implantation fluences, technologically tempered at 1000 °C (1 min) the $1\times10^{17}$ cm$^{-2}$ and as-sintered KUVI-SiO$_2$ host-matrix (reference). An intensity of all spectra was normalized to that for reference KUVI-SiO$_2$ in order to detect clearly the transformations in initial KUVI-SiO$_2$ host.

XPS O 1s core-level spectra of our samples contingently divided into two separate groups with respect to the energy location of their main peak – the 530 eV and 532.8 eV sets. The 532.8 eV set of XPS O1s spectra has no visual line-shape and binding energy (BE) deviations from the spectrum of *as-sintered* KUVI-SiO$_2$ matrix except $5\times10^{17}$ cm$^{-2}$ Re-implanted sample (Fig. 4). The latter is notably broader but still keeping the same BE-location. The local range of binding energies around 532.8 eV usually belongs to O 1s XPS lines of amorphous and crystalline SiO$_2$ polymorphs, which might be distinguished between them by either a slight BE-shift or by the dissimilar full-width-at-a-half-maximum [25, 26]. Previously we had proved this point of view in our XPS study of *a*-SiO$_2$:Sn system [25]. Thus, referencing to International XPS Databases [25, 26] and our previous XPS data [27], the established dissimilarity in FWHM is a sign of increased oxygen sublattice distortion in $5\times10^{17}$ cm$^{-2}$ Re-implanted sample, comparing to initially amorphous and *as-sintered* (not implanted) KUVI-SiO$_2$ matrix. This conclusion well agrees with the above detected and reported broadening in Re 4f core-levels of our samples, allowing to state that we have really detected the MRO-broadening at least for the implantation mode with the maximum fluence of $5\times10^{17}$ cm$^{-2}$. The sample being Re-implanted under the softest mode among applied (viz. $5\times10^{16}$ cm$^{-2}$) is free from this broadening and the identity of O 1s line with that for reference *as-sintered* KUVI-SiO$_2$ indicates, that the oxygen sublattice was not changed notably from initial tetrahedral configuration.

Another situation arises with $1\times10^{17}$ cm$^{-2}$ Re-implanted KUVI-SiO$_2$ (see Fig. 3). Here one can see both an essential line-shape and BE-location dissimilarity with SiO$_2$-like set of spectra at 532.8 eV. According to international NIST XPS Database [28] and Avantage Software XPS Database™ (it is a bundle package for XPS Thermo Scientific™ *K*-Alpha+™), the main line at 530 eV belongs to O 1s core-level of ReO$_2$, which might be formed by Re-ion implantation stimulated synthesis in the volume of KUVI-SiO$_2$ host-matrix. The strongest transformation of oxygen sublattice, caused by $1\times10^{17}$ cm$^{-2}$ fluence mode and aroused in spectra as ~ 3 eV binding energy shift as well as line-shape variation, is supposed. This type of atomic structure re-arrangement might occur only in the case of direct isovalent Re$^{4+}$ → Si$^{4+}$ substitution, which translocates the silicon atoms from their initial atomic positions into interstitials due to ballistic collisions. These interactions occur in such way, that ReO$_2$-phase becomes dominating, and, probably, even starts to play the glass-forming role like in lead-silica glasses (i.e. this new fabricated phase is also amorphous). In this case some residuals of SiO$_2$-phase have to be remained, performing much lower contribution into overall electronic structure of $1\times10^{17}$ cm$^{-2}$ Re-implanted sample. So the weak intensity and highly broadened band at ~ 533 eV in the O 1s core-level spectrum might be a sign of these SiO$_2$-phase traces, which, perhaps, are represented by scattered and uncoordinated against each other fragments of silicon-oxygen glassy network. In order to check-in our supposition, we specially made fast technological tempering of $1\times10^{17}$ cm$^{-2}$ Re-implanted sample at 1000 ºC (1 min.) with posterior recording of XPS O 1s (this tempering mode is usually employed for defects annihilation in SiO$_2$ glassy network). The result of tempering displays itself quite notable – one can see an essential narrowing in FWHM of ~ 533 eV band and clear increasing of XPS intensity in the SiO$_2$ local O1s range of BE's (see Fig. 3). Thus we have the basis for conclusion that ~ 533 eV band is really of SiO$_2$-like O 1s origin, reflecting the presence of only traces of initial SiO$_2$-phase after Re-ion embedding by pulsed implantation in the $1\times10^{17}$ cm$^{-2}$ mode.

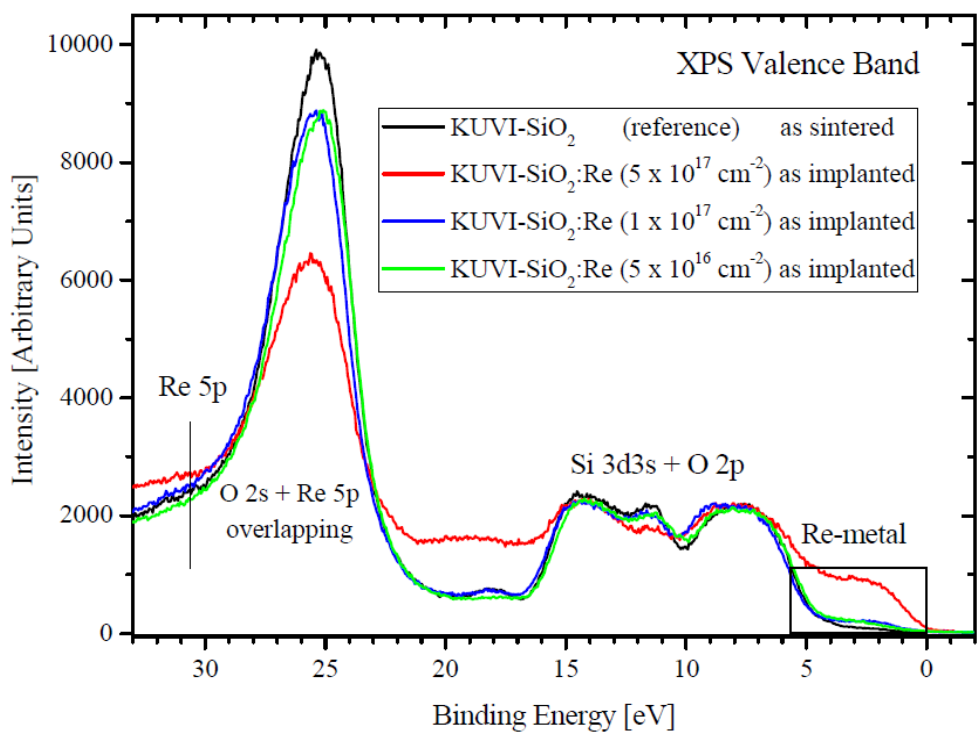

**Figure 5.** XPS Valence band (VB) mapping for KUVI-SiO$_2$:Re implanted under dissimilar fluences and *as-sintered* (not implanted) KUVI-SiO$_2$ host-matrix.

The results of X-ray photoelectron VB-mapping for our samples are presented at Fig. 5. The interpretation of partial densities of states contribution to the valence band structure of reference SiO$_2$ was made by us previously, using Thermo Scientific™ Escalab 250 Xi XPS Microprobe™ with the direct overlay of valence electron transition X-ray emission spectra of partial components – silicon and oxygen [28] (this method had been firstly implemented by X-ray Spectroscopy Laboratory, Russian Academy of Sciences – Ural Branch, Institute of Metal Physics, Yekaterinburg, Russia, see Refs. [22, 29-31]). The XPS VB spectral data, obtained for reference SiO$_2$ matrices with the help of previously employed Thermo Scientific™ XPS Microprobe spectrometer and Thermo Scientific™ *K*-Alpha+™ are almost identical, proving the high-reliability of performed XPS experiments. In particular, one can see the same dive-down effect on O 2s electronic states intensity for implanted samples as well as the simultaneous transformation of Si 3d3s – O 2p hybridized states, as it had been established for the first time and explained in [26], but now our spectra are a bit wider due another implantation modes and longer time-range of ion-beam treatment. This is causing the higher continual MRO-disordering (see the discussion above). Re 5p electronic states are usually located in the area of ~ 30–31 eV, thus partially overlapping with O 2s band and this might also contribute additionally to the broadening of O 2s band in the fabricated by ion-beam treatment VB structure [25, 26]. Al least for $5\times10^{17}$ cm$^{-2}$ sample some additional weak intensity is seen in this BE-range (Fig. 4).

The most new and notable spectral feature, appeared at the vicinity of $E_F$, (the region from 0 up to ~ 5 eV) is the broad shoulder, being caused with a high degree of probability by Re-metal contribution. This shoulder is present in the VB spectrum of $5 \times 10^{17}$ cm$^{-2}$ KUVI-SiO$_2$:Re sample, viz the sample with the maximum implantation mode. Here Re-metal phase was established using XPS core-level analysis, and pretty much absent in the VB's structure of the samples modified by softer implantation modes (see Fig. 5). The relatively low contribution of Re-metal XPS signal into Re 4f core-level spectra for these samples does not contradict the offered interpretation for the appearance of wide-and-broad shoulder located at the vicinity of $E_F$ and well coincides with the fluence dependence of spectral features reported and discussed above. We suppose that rhenium oxide phases are also contributing to the VB structure of the samples under study but actually their contribution is not essential even in core-levels (see Fig.2), so the overall origin of the valence band remains as SiO$_2$-like. Another probable reason for SiO$_2$-like valence band structure might be isovalent Re$^{4+}$ → Si$^{4+}$ substitution occurred in the glassy network in such way, that the oxygen neiboiring remains tetrahedral, as in initial SiO$_2$. In this case the key Re–O bonding distances have to be very close to that for original SiO$_4$ tetrahedra, but it seems incredibly due to the strongly dissimilar electronic configurations for Re and Si atoms from the very beginning and, hence, dissimilar Re$^{4+}$ and Si$^{4+}$ ionic radii. As an alternative, the isovalent substitution with the involvement of neutral oxygen vacancy might occur. This neutral oxygen vacancy is well-known and usually represents the irregular ≡Si–Si≡ bonding in SiO$_2$-host, being responsible for the MRO-distortions in the silica-oxygen Zachariasen-Warren three-dimensional glassy network. Pro forma, this type of defect might be considered as an Oxygen Deficient Center (ODC) [32, 33] which is well-detected by a luminescence spectroscopy technique (will be the subject of our posterior separate and extended study). Returning back to the discussion of Re-embedding scenarios into KUVI-SiO$_2$, one can suppose the [Re$^{4+}$ + vO$^0$ (≡Si–Si≡)] → Si$^{4+}$ mechanism of quazi-isovalence substitution. This means that the initial ODC's will be charge-modified by rhenium-ion localization in their vicinity as ≡Si–Si≡ ... Re$^{4+}$ and will be highly sensitive to the tempering treatment. This assumption does not contradict our XPS O 1s data (see Fig. 3, red and blue spectra) which is pointing out the sensitivity of O 1s core-levels to annealing procedure. Recall, that an isovalence substitution might successfully occurs if the principle of crystal- chemical compliance is performed – the radii difference doesn't exceed the 15 % critical value.

Opposite to isovalence substitution the non-isovalence mechanisms can be involved. They are: (i) the initiation of the charged oxygen vacancy coupled with rhenium ion [Re$^{2+}$ + vO$^{2+}$] → Si$^{4+}$ and (ii) the initiation of $E'$-defect center coupled with rhenium ion [Re$^{3+}$ + {vO$^+$ (≡ Si· Si ≡)}] → Si$^{4+}$. Here {vO$^+$ (≡ Si· Si ≡)} denotes this $E'$-defect and the dot-symbol denotes the dangling bond. Since the latter mechanism of non-isovalence substitution seems to us not enough realistic opposite to the first one (no stable at normal ambient rhenium oxide phases with $Re^{3+}$ valence state are known) so the discussion of this mechanism will be omitted.

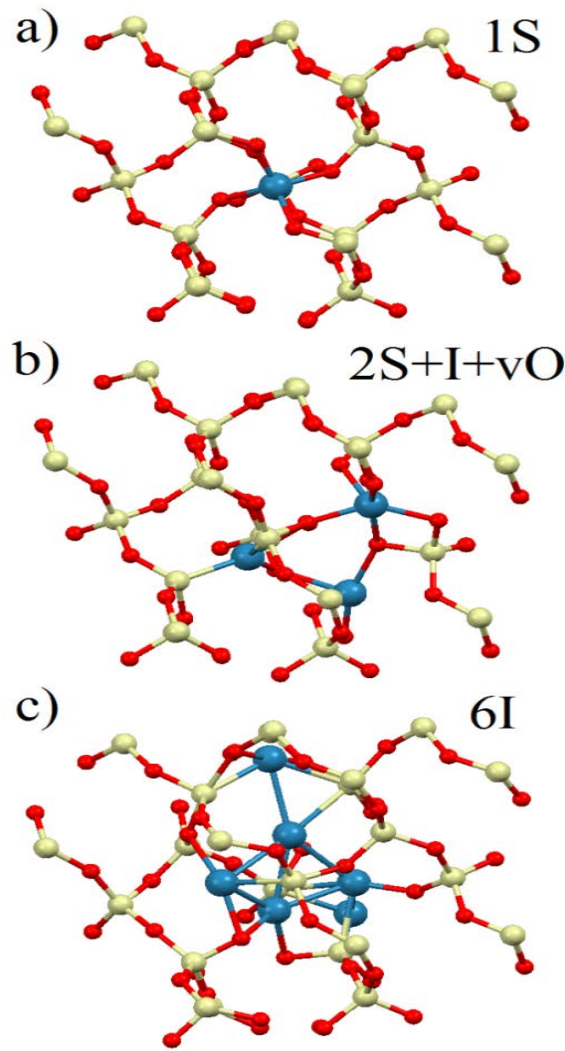

**Figure 6.** An optimized atomic structure of several configurations for substitutional (S) and interstitial (I) Re-defects and oxygen vacancies (vO).

For a detailed description on atomic level the structure fabrication scenarios for Re-defects we had calculated various atomic and electronic configurations as well as their formation energies ($E_{form}$) in $SiO_2$-host. We have to note that calculations are depicting the tendencies only at the initial stages of formation the larger configuration of defects. Based on our previous modeling experience of $SiO_2$:Zn system [12], now we also took into account not only substitutional defects (nS, Fig. 5a) but (i) the substitutional and interstitial defects (nS+I, Fig. 5b), (ii) large amount of interstitial defects without substitutional ones (nI, Fig. 5c) and (iii) the host-ions displacement into interstitial void after substitutional defect formation (1S+I(Si)). The presence of unavoidable oxygen vacancies (vO) in $a$-SiO$_2$ matrix has been taken into account as well. From materials science point of view, the most common and relatively well-studied types of defects in silicon dioxide are classified now as follows: (i) non-bridging oxygen compensated with an embedded cation (NBO-defect), (ii) doubly linked NBO-defect (peroxide bridge), (iii) simple oxygen vacancy (silicon bridge), (iv) peroxide radical (neutral NBO), (v) three-coordinated silicon atom ($E$ '- defect) and (vi) oxygen deficient clusters (ODC). That is why an employment of oxygen subnetwork imperfections into DFT-modeling of atomic and electronic structure seems as a justified key-point for the valid theoretical description of implantation-induced transformations of glassy materials under study.

**Table I.** Formation energies (in eV/defect) for various configurations of substitutional (S) and interstitial (I) Re-defects with and without oxygen vacancies (vO). The energetically most favorable configurations are bolded.

| Configuration of defect | "pure" | +vO |
|---|---|---|
| 1S | 7.00 | 5.46 |
| 1S+I(Si) | 6.80 | 5.78 |
| 2S | 6.43 | 5.85 |
| I | 5.47 | 5.66 |
| S+I | 7.33 | **4.58** |
| 2S+I | 5.09 | **4.41** |
| 2I | 4.88 | 4.90 |
| 3I | 5.03 | 5.12 |
| 4I | **4.37** | 4.90 |
| 5I | **4.44** | 5.03 |
| 6I | **4.31** | 4.75 |

The $E_{form}$ data in Table I surely demonstrate that incorporation of single Re-ion into substitutional position (1S) requires rather large values of formation energies (+7 eV). The reason for that consists in about two times difference in ionic radii of the "host" and "guest" ions which will provide even visible distortions in the vicinity of defect (Fig. 6a). The presence of oxygen vacancy in glassy network is vastly responsible for the initialization of distortion process of SiO$_2$-matrix structure by means of releasing more inner space for larger Re-ion embedding. Depending upon the type of oxygen vacancy occurred (see the classification above), the formation energy might be reduced and another process will be dramatically preferable. The same tendency is also valid for the pair of substitutional defects (2S).

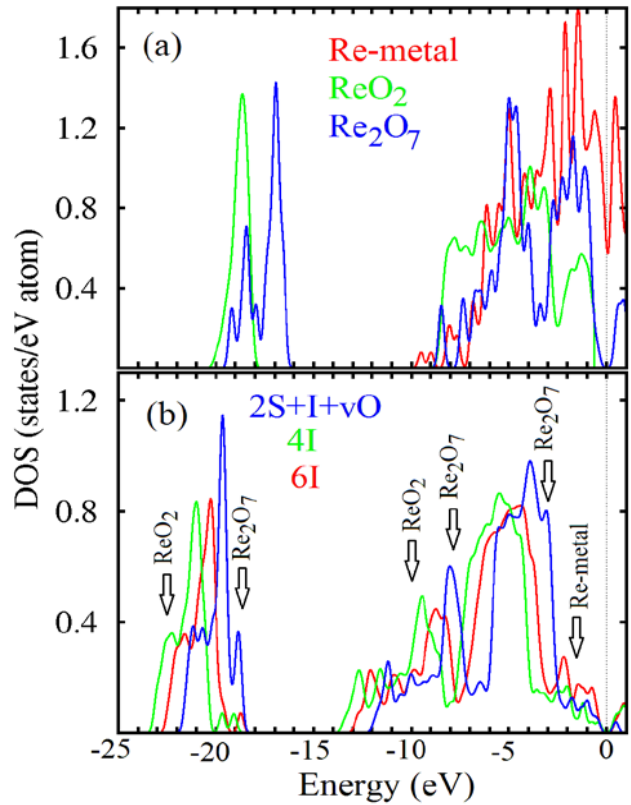

**Figure 7.** Total densities of states (DOSes): (a) for Re-metal and rhenium oxides; (b) for *a*-SiO$_2$:Re with the most energetically favorable configurations of defects (see Table I for details). Arrows indicate the specific DOS-features corresponding to local structures similar to that shown on (a).

In contrast with Zn-incorporation into SiO$_2$-matrix [12], the displacement of Si-ion to the interstitial void as a result of Re-implantation does not change the energetics valuably. The combinations of substitutional and interstitial defects (S+I and 2S+I) require formation energies of almost the same order, but the presence of oxygen vacancies will reduce formation energies by about 2.5 eV, which does not seem insignificant. In the case of only interstitial defects (nI) the calculations of formation energies demonstrate the tends to decrease the $E_{form}$ with a light alteration in the energy values for even and odd number of interstitial impurities. For the number of interstitial defects above 4, the formation energies reached their minimal values (viz less than 4.5 eV) and became of the same order as in S+I+vO and 2S+I+vO defects. Note, that the presence of oxygen vacancies increases the formation energies for nI-configurations. Thus the results of our calculations demonstrate that the two dissimilar patterns of defects distribution have nearly almost the same energies. One type of mixed defects (nS+I+vO, n=1,2) might be realized in the case of Re-impurity low concentrations (low doping mode) and only in combination with oxygen vacancies; the second type (nI, n>3) requires larger impurity concentration and absence of any type of oxygen vacancy as well.

In order to compare presented above our theoretical considerations with the XPS core-level and VB measurements we have combined and plotted calculated Densities of States for the most energetically favorable configurations of defects (see Table I). It should be noted that the electronic structure for S+I+vO configuration which is almost the same as for 2S+I+vO is omitted on Fig. 7 for clarity. For an identification of similarity with various Re-based compounds we also performed the calculation of electronic structure of Re in a pure metallic phase, and two wider spread and stable rhenium oxides – $ReO_2$ and $Re_2O_7$ (Fig. 7a). The results of calculations shown at Fig. 6b demonstrate that in the case of nS+I+vO defect configuration the final electronic structures contain typical features of $Re_2O_7$. This highest rhenium oxide contains rhenium ions both in tetrahedral and octahedral oxygen environment, having quite complicated atomic structure which on the main base is similar to the nS+I+vO configurations (Fig. 7b). For the case of 4I configuration some features of Re-metal appear, but $ReO_2$-phase traces are still prevailing. In contrast with $Re_2O_7$, rhenium dioxide has classic rutile-like structure – viz rhenium here affords only octahedral oxygen environment. Transferring this scenario into $a$-$SiO_2$:Re might be possible solely for substitutional defects. An extensive appearance of Re–Re bonds is responsible for some DOS-contribution from metallic Re. The onward increasing of interstitial impurities will increase the contribution of Re–Re bonds and, as a result, will transform the electronic structure into the structure with a metal-like character. No discrepancy with experimentally obtained XPS data on the VB structures was found in our theoretical consideration.

The comparison of energies and calculated electronic structure of various configurations of Re and oxygen defects allows to draw the following scenarios of Re-embedding into $a$-$SiO_2$-host: (i) at low Re-impurity concentrations in the vicinity of oxygen vacancies the formation of nS+I configurations occurs which highly corresponding to the $Re_2O_7$-like atomic structure and reciprocal contributions in XPS spectra; (ii) in the areas with high Re-impurity concentrations the interstitial Re-clusters are formed (nI, n<5) corresponding to $ReO_2$-like and Re-metal atomic structure and reciprocal contributions in XPS spectra; (iii) the onward increasing of impurity concentration provides the appearance of precursors and growth of nI-clusters with an amplification of Re-metal contribution majority to the final VB structure. Finally, the MRO-distortions in initial oxygen network of the $a$-$SiO_2$ host-matrix because of the appeared mixed defect configurations being induced by Re-embedding were surely established by XPS and DFT combined study.

## Conclusions

The combined XPS-and-DFT study of KUVI-SiO$_2$:Re (type IV) with referencing to the latest accepted view-points on structural imperfections in *a*-SiO$_2$ allows to derive the following majority in our experimental and theoretical data. The maximum $5\times10^{17}$ cm$^{-2}$ Re-implantation mode oversaturates the host-matrix with Re-atoms in such a way that even separate Re-metal phase appears and accumulates as interstitials. This accumulation strongly transforms the vicinity band-tail states in VB, causing the reduction of energy band-gap and the strongest MRO-distortions as compare with as-sintered KUVI-SiO$_2$ (Re-metallization mode of pulse Re-implantation). The lower modes of implantation results in ion-stimulated synthesis of prevailing and most stable ReO$_2$-phase in the volume of host-matrix, which means that formally Re$^{4+}$ → Si$^{4+}$ substitution occurred with a strong re-arrangement and further amorphyzation of oxygen network which in common becomes similar to ReO$_2$. As an additional effect the growth of Re$_2$O$_7$-like atomic structure was clearly established. The discussed scenarios might be of practical interest for the functionalization of KUVI-SiO$_2$ hosts or used as an intermediate pre-form for the onward technological separation with other technological methods the only Re-metal phase in the volume of employed matrix, only ReO$_2$ or/and Re$_2$O$_7$.

**Acknowledgements**

The synthesis of KUVI-SiO$_2$ samples and posterior Re ion-implantation treatment were supported by the Act 211 of the Government of the Russian Federation (Agreement No. 02.A03.21.0006) and the Government Assignment of Russian Ministry of Education and Science (Contract No. 3.2016.2014/K). Technical support in the XPS measurements of the studied samples and the XPS Thermo Scientific™ *K*-Alpha+™ spectrometer provided by the Ural Center for Shared Use "Modern Nanotechnology" (Ural Federal University, Yekaterinburg, Russia) are gratefully acknowledged.

**References**

[1] S. Lee, *Encyclopedia of Chemical Processing*, Ed.: CRC Press, ISBN 0824755634, 2006.

[2] R. Doering, Y. Nishi, *Handbook of Semiconductors Manufacturing Technology*, Ed.: CRC Press, ISBN 1574446754, 2007.

[3] M.-L. Liu, D.-A. Yang, Y.-F. Qu, J. Non.-Cryst. Solids 354 (2008) 4927.

[4] B. Mahltig, H. Bottcher, J. of Sol-Gel Sci. Tech. 27 (2002) 43.

[5] Th. Demuth, Y. Jeanvoine, J. Hafner, J.G. Ángyán, J. Phys.: Condens. Matter 11 (1999) 3833.

[6] S.-N. Luo, O. Tschaune, P.D. Asimow, T.J. Ahrens, Am. Mineralogist 89 455 (2004).

[7] D.A. Zatsepin, A.F. Zatsepin, D.W. Boukhvalov, E.Z. Kurmaev, N.V. Gavrilov, N.A. Skorikov, A. von Czarnowski, H.-J. Fitting, Phys. Stat.Solidi B 252 (2015) 2185.

[8] I. Sunagawa, H. Iwasaki, F. Iwasaki, *Growth and Morphology of Quartz Crystals: Natural and Synthetic*, Ed.: TerraPub, Tokyo, ISBN 978-4-88704-146-2, 2009.

[9] *Materials Science with Ions Beams*, Edited by Dr. H. Bernas, Ed.: Springer-Verlag Berlin Heidelberg, ISBN 978-3-540-88789-8 (eBook), 2010.

[10] B. Schmidt, K. Wetzig, *Ch. 4. Materials Processing*, In the book: *Ion Beams in Materials Processing and Analysis*, Authors: B. Schmidt, K. Wetzig, Ed.: Springer-Verlag Wien, ISBN 978-3-211-99356-9 (eBook), 2013.

[11] C.R. Hammond, *The Elements. Handbook of Chemistry and Physics*, Ed.: CRC Press (The 81$^{st}$ edition), ISBN 0-8493-0485-7, 2004.

[12] D.A. Zatsepin, D.W. Boukhvalov, E.Z. Kurmaev, N.V. Gavrilov, S.S. Kim, I.S. Zhidkov, Appl. Surf. Science 379 (2016) 223.

[13] J.F. Ziegler, J.P. Biersack, M.D. Ziegler, SRIM – The Stopping and Range of Ions in Matter, Ion Implantation Press (2010). Electronic manual, available at: http://dtic.mil/dtic/tr/fulltext/u2/a515302.pdf

[14] W.B. Kim, M. Nishiyama, H. Kobayashi, J. Electron Spectroscopy and Rel. Phenomena 176 (2010) 8.

[15] J. M. Soler, E. Artacho, J. D. Gale, A. Garsia, J. Junquera, P. Orejon, D. Sanchez-Portal, J. Phys.: Condens. Matter., 14 (2002) 2745.

[16] J. P. Perdew, K. Burke, M. Ernzerhof, Phys. Rev. Lett. 77 (1996) 3865.

[17] O.F. Sankey, D.J. Niklewski Phys. Rev. B, 40 (1989), 3979.

[18] N. Troullier, J.L. Martins Phys. Rev. B, 43 (1991), 1993.

[19] E. Deligoz, K. Colakoglbu, H. B. Ozisik, Y. O. Ciftci, Solid State Comm. 151 (2011) 1122.

[20] Y. Liang, X. Yuan, W. Zhang, J. Appl. Phys. 109 (2011) 053501.

[21] T. Alonso-Lana, A. Ayuela, F. Aguilera-Granja, Phys. CVhem. Chem. Phys. 18 (2016) 21913.

[22] V.R. Galakhov, L.D. Finkelstein, D.A. Zatsepin, E.Z. Kurmaev, A.A. Samokhvalov, S.V. Naumov, G.K. Tatarinova, M. Demetr, S. Bartkowski, M.Neumann, A. Moewes, Phys. Rev. B 62 (2000) 4922.

[23] X. Ou, et. al., ACS Appl. Mater. Interfaces 23 (2013) 12764.

[24] S. Hashimoto, A. Tanaka, Surface and Interface Analysis 34 (2002) 262.

[25] Thermo Scientific XPS: Knowledge Base, © 2013, http://xpssimplified.com/knowledgebase.php (called 2016-05-30).

[26] NIST XPS Database, version 4.1, http://srdata.nist.gov/xps/ (called 2016-05-30).

[27] D.A. Zatsepin, A.F. Zatsepin, D.W. Boukvalov, E.Z. Kurmaev, N.V. Gavrilov, Appl. Surf. Sci. 367 (2016) 320.

[28] D.A. Zatsepin, P. Mack, A.E. Wright, B. Schmidt, H.-J. Fitting, Phys. Stat. Sol. A 208 (2011) 1658.

[29] E.Z. Kurmaev, V.R. Galakhov, D.A. Zatsepin, V.A. Trofimova, S. Stadler, D.L. Ederer, A. Moewes, M.M. Grush, T.A. Callcott, J. Matsuno, A. Fujimori, S. Nagata, Solid State Comm. 108 (1998) 235.

[30] D.A. Zatsepin, V.R. Galakhov, M.A. Korotin, V.V. Fedorenko, E.Z. Kurmaev, S. Bartkowski, M. Neumann, R. Berger, Phys. Rev. B 57 (1998) 4377.

[31] E.Z. Kurmaev, A.V. Postnikov, H.M. Palmer, C. Greaves, St. Bartkowski, V. Tsurkan, M. Demetr, D. Hartmann, M. Neumann, D.A. Zatsepin, V.R. Galakhov, S.N. Shamin, V.A. Trofimova, J. Phys.: Condens. Matter 12 (2000) 5411.

[32] L. Skuja, J. Non-Crystalline Solids, 239 (1998) 16.

[33] A.F. Zatsepin, E.A. Buntov, A. Pustovarov, H.-J. Fitting. J. of Luminescence 143 (2013) 498.